\documentclass[12pt,prd,aps,floatfix]{revtex4-1}
\pdfoutput=1
\usepackage{amsmath}
\usepackage{graphicx}
\usepackage{graphics}
\usepackage{mathrsfs}
\usepackage{hyperref}
\usepackage{color}
\usepackage[T1]{fontenc}
\usepackage[latin9]{inputenc}
\usepackage{geometry}
\usepackage{bookmark}
\usepackage{array}
\usepackage{amssymb}
\usepackage{cancel}
\usepackage{graphicx}
\usepackage{esint}
\usepackage{longtable}
\usepackage{multirow}
\usepackage{amsmath}
\usepackage{amssymb}
\usepackage{cancel}

\providecommand{\\}{\\}

\newcommand{\bea}{\begin{eqnarray}}
\newcommand{\eea}{\end{eqnarray}}
\newcommand{\beq}{\begin{equation}}
\newcommand{\eeq}{\end{equation}}

\def\slash#1{\mbox{$\not\!\! #1$}}

\def\simge{\mathrel{\rlap{\raise 0.511ex \hbox{$>$}}{\lower 0.511ex
 \hbox{$\sim$}}}}
\def\simle{\mathrel{\rlap{\raise 0.511ex \hbox{$<$}}{\lower 0.511ex
 \hbox{$\sim$}}}}
\def\slash#1{\setbox0=\hbox{$#1$}\dimen0=\wd0 \setbox1=\hbox{/} \dimen1=\wd1
 \ifdim\dimen0>\dimen1 \rlap{\hbox to \dimen0{\hfil/\hfil}} #1
 \else \rlap{\hbox to \dimen1{\hfil$#1$\hfil}} / \fi}

\begin{document}

\title{Metadynamics Surfing on Topology Barriers: the
  \texorpdfstring{$CP^{N-1}$}{CP(N-1)} Case} \author{A.~Laio}
\affiliation{SISSA, Via Bonomea 265, I-34136, Trieste,}
\author{G.~Martinelli}
\affiliation{SISSA, Via Bonomea 265, I-34136, Trieste, and INFN
  Sezione di Roma La Sapienza Piazzale Aldo Moro 5, 00185 Roma,
  Italy,}
\author{F.~Sanfilippo}
\affiliation{School of Physics and Astronomy, University of
  Southampton, Southampton SO17 1BJ, UK.}

\begin{abstract}
  \vskip 0.5cm As one approaches the continuum limit, $QCD$ systems,
  investigated via numerical simulations, remain trapped in sectors of
  field space with fixed topological charge. As a consequence the
  numerical studies of physical quantities may give biased results.
  The same is true in the case of two dimensional $CP^{N-1}$ models.
  In this paper we show that {\it metadynamics}, when used to simulate
  $CP^{N-1}$, allows to address efficiently this problem. By studying
  $CP^{20}$ we show that we are able to reconstruct the free energy of
  the topological charge $F(Q)$ and compute the topological
  susceptibility as a function of the coupling and of the volume. This
  is a very important physical quantity in studies of the dynamics of
  the $\theta$ vacuum and of the axion. This method can in principle
  be extended to $QCD$ applications.
  
\end{abstract}
\maketitle
\section{Introduction}
\label{intro}

In numerical simulations of asymptotically free theories, $CP^{N-1}$
in two dimensions and $SU(N)$ in four dimensions, one observes an
increase of the autocorrelation time $\tau$, defined as the number of
iterations needed to generate independent field configurations, as one
proceeds towards the continuum limit where the coupling constant
vanishes and the typical correlation length diverges. This corresponds
to the critical slowing down occurring in statistical systems close to
a second order phase transition. In addition, in these systems, a
particularly dramatic increase of autocorrelation time is observed in
the case of the topological charge, independently of the precise
discretized definition which is used in the simulation on the
lattice~\cite{Rossi:1993nz, DelDebbio:2002xa, DelDebbio:2004xh,
  Luscher:2010we, Schaefer:2010hu, Luscher:2011kk, Chowdhury:2013mea,
  Brower:2014bqa, Flynn:2015uma, Engel:2011re}.

In general, the asymptotic scaling behavior of the autocorrelation
time with the lattice spacing is expected to be power-like for all the
quantities. On the contrary, in both $CP^{N-1}$ and $QCD$ the
autocorrelation time of the topological charge, $\tau_Q$, grows so
rapidly with the length scale $\xi$, that an apparent exponential
behavior $\tau_Q \sim \exp(c\, \xi^\theta)$ in the explored range of
values of $\xi$ is compatible with the available
data~\cite{DelDebbio:2002xa, DelDebbio:2004xh}. On the other hand,
this peculiar effect is not observed for ``quasi-Gaussian modes'' such
as the plaquette or the Polyakov line correlations, suggesting a
separation of the dynamics of the topological modes from that of
quasi-Gaussian ones.

The different dynamical behavior of quasi-Gaussian and topological
modes is induced by sizable free-energy barriers separating different
regions of the configuration space. Therefore the evolution in this
space presents a long relaxation time due to the transitions between
different topological charge sectors, and the corresponding
autocorrelation time is expected to behave as $\tau_Q\sim \exp(\Delta
F(Q))$, where $\Delta F(Q)$ is the typical free-energy barrier between
different topological sectors. If the height of the barrier increases
as we proceed toward the continuum limit, the system can be trapped in
a topological-charge sector for a number of simulation steps
comparable or even longer than the total available resources of the
simulation. In $QCD$ this problem may bias the lattice predictions for
several important physical quantities such as the mass of the
$\eta^\prime$ meson~\cite{Veneziano, Witten, DelDebbio:2004ns,
  Ce:2015qha}, or the polarized baryon structure
function~\cite{Shore:2007yn, Aidala:2012mv, Marchand, Mulders}. In
general this bias will be present for any observable correlated to the
topological charge, whenever the algorithm does not explore with the
appropriate weight the different topological sectors. This has been
shown to occur in ref.~\cite{Flynn:2015uma} in the contest of
$CP^{N-1}$ models. It is clear that in order to design a cure for
these problems one should also measure and understand how the typical
free-energy barriers scale with the correlation length (and with the
physical volume) in a simulation.

In the recent literature the problem of very long autocorrelation
times and of the theoretical control over the systematic error in
numerical simulations has been addressed in a series of
papers~\cite{DelDebbio:2002xa, DelDebbio:2004xh, Schaefer:2010hu,
  Engel:2011re, Chowdhury:2013mea, Brower:2014bqa, Flynn:2015uma} and
several solutions to the topological critical slowing down have been
proposed~\cite{deForcrand:1997fm, Luscher:2010we, Luscher:2011kk,
  Ramos:2012bb, McGlynn:2013ava, Gerber:2014bia, Endres:2015yca,
  Gambhir:2015rha, Dromard:2015nba}.

In this paper we propose to address the problem of topological
trapping by using {\it metadynamics}, a powerful method that was
introduced to enhance the probability of observing rare conformational
changes and reconstructing the free energy in biophysics, chemistry
and material sciences systems~\cite{laio2002escaping,laio2}.

The metadynamics method~\cite{laio2002escaping,laio2} requires the
preliminary identification of Collective Variables (CV) which are
assumed to be able to describe the process of interest. A CV is a
(physical) quantity which depends on a set of the coordinates or
fields of the system. In the simplest case the variable could be just
the topological charge itself although the full power of this approach
relies in its ability to treat several CVs simultaneously. The
dynamics in the space of the chosen CVs is enhanced by a
history-dependent potential constructed as a sum of Gaussians centered
along the trajectory followed by the CVs. The sum of Gaussians is
exploited to reconstruct iteratively an estimator of the free
energy. This procedure resembles the Wang and Landau
algorithm~\cite{wang2001}, in which a time-dependent bias is
introduced to modify the density of states to produce flat histograms
in models with discrete energy levels.

The history of microscopic systems in normal Monte Carlo or Hybrid
Monte Carlo (HMC)~\cite{Engel:2011re} corresponds to a random walk
with a bias in the direction of lower free energy. In systems with
many local minima the probability to explore the transition regions
and tunnel in a different minimum is very small, particularly when the
height of the barrier is high. In metadynamics, the system has access
to a feedback which during the time evolution fills the local free
energy minima. Thus, even if at the beginning the system visits more
often the region at the bottom of a local minimum, after a few steps
almost deterministically it starts exploring regions corresponding to
higher and higher values of the free energy. Sooner or later, the
system fills the minimum, climbs out of it, and visits another minimum
that is eventually also filled, until all the relevant minima are
explored.
The key idea of metadynamics is exploiting the time-dependent bias
potential itself as a free energy estimator. In particular, the time
average of the bias potential has been shown to converge to the
negative of $F$ with an error that scales to zero with the inverse
square root of the simulation time~\cite{laiostar}.

We here propose to address the problem of topological trapping by
performing metadynamics on the topological charge. We will show that
this approach induces a large number of transitions between different
sectors, and therefore converges very rapidly. At the same time, the
approach allows computing the unbiased average value of any observable
by standard reweighting techniques.
In order to test our proposal we first study the two-dimensional
$CP^{N-1}$ models that have several features in common with $QCD$,
such as asymptotic freedom and a non-trivial topological
structure. Since these models require much smaller computing
resources, they are an ideal theoretical laboratory to be used in an
early and exploratory stage of any new algorithm. We find the
improvement induced by metadynamics considerable and worth to be
implemented in a $QCD$ study that we plan to perform in the near
future.

The paper is organized as follows: in Sec.~\ref{sec:cpn} we recall the
basic ingredients of $CP^{N-1}$ and define the quantities that will be
measured in our numerical simulation; in Sec.~\ref{sec:meta} the
metadynamics algorithm is described; in Sec.~\ref{sec:nume} we present
the results of our numerical study and compare several quantities
obtained without or with metadynamics. Our conclusions will be given
in Sec.~\ref{sec:conclu} and some technical details are presented in
appendix~\ref{app:SmeU1}.

\section{\texorpdfstring{$CP^{N-1}$}{CP(N-1)} and the different definitions
  of the topological charge on the lattice}
\label{sec:cpn}
In the continuum the two dimensional $CP^{N-1}$ model is defined by
the action \bea S= \frac{1}{g} \, \int \, d^2x \, {\bar D_\mu \bar z}
D_\mu z \, , \eea where $z$ is a complex $N$-dimensional field with
unit norm $\bar z \cdot z = \sum_{i=1}^N \, z^*_i z_i =1$ and the
covariant derivative is given by $D_\mu = \partial_\mu + i A_\mu$.
The field $A_\mu$ has no kinetic term. For this reason, by using the
equation of motion, $A_\mu$ can be expressed in terms of the field $z$
\bea
A_\mu = \frac{i}{2} \left( \bar z \cdot \partial_\mu z - \partial_\mu
  \bar z \cdot z \right) \, .
\eea
The action is invariant under the local $U(1)$ gauge transformation
\bea
z(x) \to e^{i \Lambda(x)} \, z(x) \, , \quad \quad A_\mu \to A_\mu
- \partial_\mu \Lambda(x) \, ,
\eea
The connected correlation function and magnetic susceptibility are
defined as
\bea
G( x) &=& \langle {\it Tr} P( x) P(0) \rangle - \frac{1}{N} \langle
{\it Tr} P(x) \rangle \langle {\it Tr} P( 0) \rangle\, , \quad \quad
P(x) \equiv \bar z(x) \otimes z(x) \quad \left( P_{i,j}(x) = z^*_i z_j
\right) \nonumber \\ \nonumber \\ \chi_m &=& \int \, d^2x \, G(x) =
\tilde G(p =0)\, .  \label{eq:proj}
\eea
In the continuum one can define the correlation length $\xi$ from the
large time-distance behavior of $G(x)$ \bea G(x_0) = \int dx_1 \,
G(x_0,x_1) \sim e^{-x_0/\xi_W} \, , \label{eq:gt} \eea or from the
second moment of the correlation function
 \bea
 \xi_{II} = \frac{\int d^2x \, \frac{x^2}{4}\, G(x) }{ \int d^2x \,
   G(x) } \, , \label{eq:gt1}
 \eea
In our numerical analysis we have used the definition $\xi_G$ given
below in eq.~(\ref{eq:xig}), which is proportional to $\xi_W$ and
$\xi_{II}$ in the scaling region.

The topological-charge density $q(x)$, the total topological charge
$Q$ and the topological susceptibility $\chi_t$ are defined as
\bea
q(x) &= &\frac{1}{2 \pi} \, \epsilon_{\mu\nu} \partial_\mu A_\nu \, ,
\quad \quad Q = \int \, d^2x \, q(x) \, , \nonumber \\ \chi_t &=& \int
\, d^2x \, \langle q(x) q(0) \rangle \ = \frac{\langle Q^2 \rangle}{V}
\, .
\eea
The lattice action is given by~\cite{DiVecchia:1981eh}
\bea
S= \frac{1}{g} \, \sum_{\vec n, \mu} \, {\bar D_\mu \bar z_{\vec n}}
D_\mu z_{\vec n} \, , \label{eq:latac}
\eea
where we introduce the lattice covariant derivative
\bea
D_\mu \bar{z}_{\vec n} = \lambda_{\vec n, \hat \mu} \bar{z}_{\vec n +
  \hat \mu}-\bar{z}_{\vec n } \, ,
\eea
expressed in term of the $U(1)$ gauge link $\lambda_{\vec n, \hat \mu}
\equiv \exp(i A_\mu(\vec n + \hat \mu a/2)$, where $a$ is the lattice
spacing and $\hat \mu$ the unit vector in the direction $\mu$.  This
is the simplest nearest neighbor derivative. Improved versions of it
could also be used. Lattice gauge invariance, in its linear
realization, reads
\bea
\lambda_{\vec n, \hat \mu} \to e^{i \Lambda(x)} \lambda_{\vec n, \hat
  \mu} e^{- i \Lambda(x+\hat \mu)}\, , \quad\quad \bar{z}_{\vec n }
\to e^{i \Lambda(x)}\bar{z}_{\vec n } \, .
\eea
By integrating by part the term in eq.~(\ref{eq:latac}) and
removing the term independent of the fields we arrive to the action
used in our numerical simulation
\bea
S= -\frac{2}{g} \, \Re [ \, \sum_{\vec n, \mu} \, \bar z_{\vec n+\hat \mu
}\lambda_{\vec n, \hat \mu} z_{\vec n} \,-1] \,
. \label{eq:latac2}
\eea
We use the lattice definition of the correlation length
$\xi_G$~\cite{DelDebbio:2004xh}
\bea
\xi_G^2 = \frac{1}{4 \sin^2(q_m/2)} \, \frac{\tilde G(0)- \tilde
  G(q_m)}{\tilde G(q_m)}\, , \label{eq:xig}
\eea
where $q_m$ is the smallest non zero dimensionless momentum on a
lattice with lattice spacing $a$ and physical volume $aL$, namely
$q_m= (2\pi/L,0)$. In the continuum limit $a \xi_G \to \xi_W$, where
the correlation $\xi_W$ was defined in eq.~(\ref{eq:gt}).

At finite lattice spacing no unique definition of the topological
charge exists. One possible geometrical definition of the topological
charge was introduced in Ref.~\cite{Berg:1981er} as

\begin{equation}
  Q^g=\frac{1}{2\pi} \sum_n \rm{Im} \left\{ 
    \ln \rm{Tr} \left[P(n+\hat\mu+\hat\nu)P(n+\hat\mu)P(n)\right] +
    \ln \rm{Tr} \left[P(n+\hat\nu)P(n+\hat\mu+\hat\nu)P(n)\right]
  \right\},\,\label{defGeoTopo}
\end{equation}

where $\mu\neq\nu$ and $P$ is the projector defined in
eq.~(\ref{eq:proj}). Thanks to the periodicity of the lattice, the
previous definition is guaranteed to be an exact integer number on
every configuration (strictly speaking this in fact holds for all
configurations except for a subset of measure zero). Different
geometrical definitions of the topological charge, however, are not
guaranteed to return the same integer number, especially on coarse
lattices. $Q^g$ can be written in terms of the phase $\theta_{\vec n,
  \hat \mu}={\it arg}\, ( \bar z_{\vec n} z_{\vec n +\hat \mu})$ as
\begin{equation}
  Q^g=\frac{1}{2\pi} \sum_n \, (\theta_{\vec n, \hat \mu}
  +\theta_{\vec n+ \hat \mu, \hat \nu} - \theta_{\vec n+ \hat \nu,
    \hat \mu}-\theta_{\vec n, \hat \nu})\label {defGeoTopo1}
\end{equation}

Being strictly integer-valued, the geometrical definition of
eq.~(\ref{defGeoTopo}) is insensible to infinitesimal continuum
deformations of the fields. As a side effect, a geometrical version
of the topological charge cannot be used to define a bias variable in
the metadynamics approach. Indeed the HMC algorithm defines a
continuous dynamics in a fictitious-time, which requires the
evaluation of the derivative of the action with respect to the local
variables. Any bias potential built in terms of the geometrical
definition of eq.~(\ref{defGeoTopo}) would give rise to a dynamics
completely insensible to the past history of the topological charge,
as long as the system remains confined in the same topological sector.
The system would only feel the effects of the bias potential when
crossing the boundaries between different topological sectors, where
the infinite force arising from the discontinuity in the potential
would break the numerical integration of the equations of motion.

Another definition of the topological charge that better serves our
scope is  given in terms of the imaginary part of the plaquette,
as illustrated in Ref.~\cite{Campostrini:1992ar}
\begin{equation}
  Q^\lambda=\frac{1}{2\pi}\sum_n \Im[\lambda_\mu(n)
    \lambda_\nu(n+\hat\mu) \bar\lambda_\mu(n+\hat\nu)
    \bar\lambda_\nu(n)]\quad\mu<\nu.
  \label{defNonGeoTopo}
\end{equation}
$Q^\lambda$ differs from the geometrical definition of
eq.~(\ref{defGeoTopo}). We have
\begin{equation}
  Q^\lambda=Z_Q \, Q^g+\eta  \, ,
\end{equation}
where $Z_Q$ is a suitable renormalization constant that can be
computed in perturbation theory and $\eta$ is a zero-average additive
ultraviolet noise depending on the particular field configuration, the
variance of which is a increasing function of the lattice volume and a
mildly dependent function of the lattice spacing.

Such noise can be reduced by computing the topological charge of
eq.~(\ref{defNonGeoTopo}) after ``smoothing'' the $\lambda$ fields
(see e.g. Ref.~\cite{Bonati:2014tqa} for a recent comparison between
cooling and Wilson flow in the contest of QCD), or by smearing the
gauge fields with procedures like APE~\cite{Albanese:1987ds} or
HYP~\cite{Hasenfratz:2001hp}.

Introducing a bias potential related to a non-geometric definition of
the topological charge may accelerate the dynamics of the ultraviolet
noise, that is expected to be connected with the degrees of freedom
ultimately related to the tunneling between different topological
sectors. As a matter of fact, it will turn out to be useful to filter
out the highest frequency components of the noise, with the purpose of
capturing those components expected to be more closely related
to the tunnelling phenomenon. To this end we make use of a modification
of the {\it stout smearing}~\cite{Morningstar:2003gk}, adapted to
treat $U(1)$ variables, according to the procedure explicitly
described in appendix~\ref{app:SmeU1}.

We stress that the smeared topological charge is exploited in our
approach only to enhance the tunneling between different minima, thus
improving the convergence of the system to thermalisation with the
possibility of averaging over all the different topological sectors.
After convergence is achieved, one can then compute the properties of
the system described by the normal CP$^{N-1}$ action by using standard
reweighting techniques. For example, one can compute the probability
of observing different values of the exact $Q$.

To conclude this section we define the fictitious-time autocorrelation
function $C_O(t)$, where $O$ is any observable (topological charge,
magnetization, $\xi$ etc.) and $t$ is the discrete simulation time
expressed in sweeps
\beq
C_O(t) = \langle (O(t) -\langle O\rangle ) (O(0) -\langle O\rangle)
\rangle \, ,
\eeq
where the averages are taken after the thermalisation of the
system. We expect that $C_O(t) \sim e^{-t/T_O}$ at large $t$, where $
T_O$ is the typical autocorrelation time of the observable $O$.

\section{Metadynamics}

\label{sec:meta} 
Let us consider a physical system described by a set of coordinates
$x$ and an action $S(x)$ that evolves under the effect of a given
dynamics that samples a equilibrium probability proportional to
$\exp(-S(x))$. We are interested in exploring the properties of the
system as a function of a CV $Q(x)$. The probability distribution of
this variable is given by
\begin{equation}
  P(q)=\int\, dx\,\exp(-S(x))\,\delta(q-Q(x))\,.\label{eq:free1}
\end{equation}
and the corresponding free energy is
$F\left(q\right)=-\log\left(P\left(q\right)\right)$. The probability
distribution and the free energy can be estimated by computing the
histogram of $q=Q(x)$ over a phase space trajectory $x(t)$:
$P(q)\sim\frac{1}{t}\int_{0}^{t}\,
dt^{\prime}\,\delta(q\left(t'\right)-q)$ where
$q\left(t\right)=Q\left(x\left(t\right)\right)$. If the system
displays metastability, namely if $P(q)$ is characterized by the
presence of two or more local maxima, separated by regions where the
probability of observing the system is small, this estimator is
affected by large statistical errors, since $q$ is normally trapped in
a region of high probability, and only rarely performs a transition to
another region. This is the case for the dynamics of a $CP^{N-1}$
model, where at small lattice spacing the system is trapped in a
specific topological sector.

In metadynamics the action $S\left(x\right)$ is modified by adding to
it a history-dependent potential of the form
\begin{equation}
  V_{G}(Q(x),t)=\sum_{\begin{array}{c}
      t'=\tau_{G},2\tau_{G},\ldots\\ t'<t
\end{array}}g\left(Q\left(x\right)-q\left(t'\right)\right)\label{eq:freeE}
\end{equation}
where $g\left(q\right)$ is a non-negative function of its argument,
that rapidly vanishes for large $\left|q\right|.$ In the original
implementation, $g(q)=w\exp\left(-\frac{q^{2}}{2\delta q^{2}}\right)$,
where $w$ and $\delta q$ are two parameters that can be used to tune
the growth speed of $V_{G}$. Thus, the metadynamics potential is a sum
of small repulsive potentials, acting on the CV, and localized on all
the configurations $q(t)$ that have already been explored during the
trajectory, up to time $t$. This potential disfavors the system from
revisiting configurations that have already been visited. If the
dynamics of $q$ is bound in a finite connected region, after a
transient the probability distribution of $q$ in this region can only
be approximately flat. Indeed, if this is not the case, by definition
the system would spend more time in a subregion of $q,$ and $V_{G}$
would grow preferentially in that region, disfavoring the system from
remaining there. Thus, deviations from the flat distribution can only
survive for a transient
time. $P(q)\exp\left(-V_{G}\left(q,t\right)\right)$ must be
approximately constant or, equivalently,
\begin{equation}
V_{G}(q,t)\sim-F(q)\,.\label{eq:free2}
\end{equation}
This equation states that in metadynamics the free energy is estimated
by the negative of the bias potential itself. More precisely, since
eq.~(\ref{eq:free2}) is valid at any time, the best estimator of the
free energy at time $t$ is given by the (large) time average of
$V_{G}$ up to time $t,$
\begin{equation}
  -F\left(q\right)\sim\overline{V_{G}(q,t)}=\frac{1}{t-t_{eq}}\int_{t_{eq}}^{t}dt'V_{G}(q,t')\label{eq:free3}
\end{equation}
The equilibration time $t_{eq}$ entering in eq.~(\ref{eq:free3}) is
the time at which the history dependent potential becomes
approximately stationary (or, equivalently, the probability
distribution as a function of $q$ becomes approximately flat). Like in
the ordinary estimates of the average value of an observable, the
exact choice of $t_{eq}$ influences only the convergence speed, but
not the final result. The difference between $-F$ and
$\overline{V_{G}}$ in eq.~(\ref{eq:free3}) decreases as the square
root of $t-t_{eq}$, with a prefactor that strongly depends on the
specific CV $q$~\cite{laiostar}.

Since the bias potential in eq.~(\ref{eq:freeE}) must be a
differentiable function of the fields, we use as CV the topological
charge $Q^{\lambda}$ defined in eq.~(\ref{defNonGeoTopo}) rather than
the ``integer\textquotedbl{} charge of eq.~(\ref{defGeoTopo}).

The probability of observing different values of the exact $Q$ is then
computed by reweighting, as discussed below.

The most efficient choice of the smoothing parameters is such that the
distribution probability of $Q^{\lambda}$ is the largest which still
allows an assignment to different integer values of the topological
charge. In other words, the distribution probabilities of
$Q^{\lambda}$ for different given topological sectors of $Q^{g}$ are
such that they do not overlap. This determines the optimal width of
$Q^{\lambda}$.

For the sake of computational efficiency, we store the
history-dependent potential on a regular grid of spacing $\delta q$;
$\left(q_{0},q_{1},\cdots,q_{n}\right)$, with $q_{i}=q_{0}+i\,\delta
q$. The use of the grid makes it possible to carry on metadynamics for
long runs at a fixed overhead per sweep (in computer time), whereas
the computer time with the naive procedure would linearly increase
with the number of sweeps.

At the beginning of the simulation we set
$V_{i}=V_{G}\left(q_{i}\right)=0$. Then, at every step, we
\begin{enumerate} 
\item compute the value of the CV $q(t)\equiv Q^{\lambda}(t)$;
\item find the grid interval $i$ where it falls
\begin{eqnarray*}
  i=\mathrm{int}\left(\frac{q\left(t\right)-q_{0}}{\delta q}+0.5\right) \, ;
\end{eqnarray*}

\item update the history dependent potential as follows
  \begin{equation}
    \begin{split}
      V_{i} & = V_{i}+w\left(1-\frac{q\left(t\right)-q_{i}}{\delta q}\right) \\
      V_{i+1} & = V_{i+1}+w\frac{q\left(t\right)-q_{i}}{\delta q} \, . 
    \end{split}
    \label{eq:parameters}
  \end{equation}
\end{enumerate}
Thus, in our implementation the function $g(q)$ entering in
eq.~(\ref{eq:free1}) is a sum of two triangular-shaped functions, one
centered on $q_{i}$, the other on $q_{i+1}$.

The force ruling the evolution of the fields $x$ is then changed by
adding to it the component deriving from the history-dependent
potential. Thus, the total force is
\begin{eqnarray}
-\nabla L\left(x\right)-\nabla V_{G}(Q(x),t)=-\nabla
L\left(x\right)-\frac{V_{i+1}+V_{i}}{\delta q}\nabla Q\left(x\right)
\, . \label{eq:23}
\end{eqnarray}

The two crucial parameters in this procedure are $w$ and $\delta q$.
For too large values of $w$ an accurate integration of the trajectory
would require an infinitesimal time step (for vanishing $w$
metadynamics becomes the standard HMC). The optimal grid spacing
$\delta q$ must be such that i) the potential wells are filled
rapidly, and this requires a large $\delta q$; ii) the free energy
$F(q)$, eqs.~(\ref{eq:free1}) and (\ref{eq:free2}), can be accurately
reconstructed, and this requires a small value of $\delta q$. Suitable
values for $w$ and $\delta q$ can be initially estimated in an
unbiased run by measuring the transition probability between two
topological sectors and the fluctuations of $Q^{\lambda}$ in a given
sector respectively~\cite{laio2}. The first will give us the height of
the barrier, and $w$ must be significantly smaller than this height,
while second give us an estimate of the width of the distribution.

In order to reach a stationary state in which the probability
distribution as a function of $q$ is flat it is necessary constraining
the dynamics of $q$ in a finite region. This can be done without loss
of generality by suppressing configurations, corresponding to large
values of $q$, that have a exponentially low probability weight. A
well established method~\cite{laio2} is given by the introduction of a
threshold value of the topological charge $Q^{thr}$: we disfavor
values of $Q^{\lambda}$ larger than the threshold by a repulsive force
which increases linearly starting from $Q^{thr}$
\begin{equation}
  V_{rest}(Q(x))= k\left(Q(x)-Q^{thr}\right)^{2}\,,\quad\vert
  Q(x)\vert\ge Q^{thr}\,. \label{eq:rest}
\end{equation}
$k$ is a new parameter which should be chosen of the same size as
$w/(2\delta q^{2})$. $Q^{thr}$ can be estimated from the value of the
topological susceptibility, choosing $Q_{thr}\gg \sqrt{\chi_Q L^2}$.

In summary five parameters have to be tuned, namely $w$, $\delta q$,
$Q^{thr}$, $k$ and the smearing parameter $\rho n$ discussed in
appendix~A. The grid on $q$ in eq.~\ref{eq:parameters} is chosen in
such a way that $q_1=-Q^{thr}$ and $q_{n-1}= +Q^{thr}$. The forces
from metadynamics in eq.~\ref{eq:23} are added only if $Q(x) \in
[-Q^{thr},+Q^{thr}]$.The history-dependent potential $V_i$ is updated
according to eq.~\ref{eq:parameters} also when $Q(x) \in [q_0,q_1]$ or
$Q(x) \in [q_{n-1},q_n]$. Note that, since the threshold $Q^{thr}$ is
chosen in such a way that the configurations that are disfavored by
the restraining potential in eq.~\ref{eq:rest} correspond to a
exponentially low probability, using a larger value of $Q^{thr}$ would
not change the estimate of the average value of any observable, except
for exponentially small corrections, much smaller than the statistical
error.

More details on how the parameters of this approach are chosen and
tuned to optimize the efficiency can be found in ref~\cite{laio2}.

\section{Numerical results}
\label{sec:nume}
In this section we present a comparison of results obtained by using
the standard HMC algorithm with those obtained by using
metadynamics. In particular we focus on the autocorrelation time
$\tau$ of several physical quantities among which the autocorrelation
of the topological charge $\tau_Q$ and its scaling properties in the
continuum limit.

We have studied $CP^{N-1}$, with $N=21$, at several values of the
coupling constant, with different physical volumes at fixed
correlation length and with different values of the correlation length
at fixed physical size. We have chosen $CP^{20}$ because it was
already extensively studied in the literature, see for example
refs.~\cite{Rossi:1993nz,DelDebbio:2004xh,Campostrini:1992ar}. These
papers made the choice of a large value of $N$ in order to enhance
finite size effects and the critical slowing down that they wanted to
study. We have taken the same value of $N$ for the same reasons and in
order to have a direct comparison with previous, quite precise data
regarding the main observables (energy, correlation length, magnetic
and topological susceptibility).

We performed molecular dynamics with a time-step $\Delta t=1$ in the
fictitious-time, integrating the equation of motion by means of the
Omelyan integrator, using 18 steps per trajectory to achieve a high
acceptance rate ($\mathcal{O}(90\%)$).

We present in Tables~\ref{tab:twoa} and~\ref{tab:twob} the input
parameters (coupling $\beta$, number of sweeps $N_s$, lattice size
$L$) and the results of the numerical simulations for the following
observables: the dimensionless correlation length $\xi_G$ defined in
eq.~(\ref{eq:xig}) and $L/\xi_G$; the average energy density $E= g
S/2V$ with the action $S$ defined in eq.~(\ref{eq:latac2}); the
magnetic susceptibility $\chi_m$ and the topological susceptibility
$\chi_Q$. We have chosen the lattice volume in order to work with
fixed physical finite volumes at the different values of $\beta$
($\xi_G$), namely at fixed $L/\xi_G$. For comparison, in
tables~\ref{tab:twoa} and~\ref{tab:twob} we present the values for
$\xi_G$, $\chi_m$ and $\chi_t$ obtained with the standard HMC and with
metadynamics, respectively.  As for the relevant parameters of the
metadynamics, see eq.~(\ref{eq:parameters}), we used $\delta q =0.05$
and varied $w$ as a function of the coupling ($w=0.025$ at
$\beta=0.65$ and $w=0.140$ at $\beta=0.65$) to compensate the increase
of the height of the barriers in the continuum limit.

The unbiased expectation value of an observable $O$ is computed
through the reweighting procedure
\bea
\langle O \rangle = \frac{\sum_i O_i \exp(-\overline{
    V(Q^\lambda_i)})}{\sum_i \exp(-\overline{ V(Q^\lambda_i)})}\, ,
\label{eq:rew}
\eea
where $\overline{V(Q^\lambda)})$ is defined in eq.~(\ref{eq:free3}).

Since this work describes the first application of metadynamics to
lattice field theory, in the next subsection we will present several
details about the numerical results, whereas in
Sec.~(\ref{ssec:scaling}) we discuss the scaling properties of the
efficiency as we proceed toward the continuum limit.

\subsection{A comparison of standard HMC and metadynamics}

\label{ssec:general}

We start by considering a metadynamics run performed at $\beta=0.70$
and $=L=62.$ For these values, the standard HMC is still able
to explore different topological sectors, and achieve convergence.
This allows checking if the two approaches provide consistent results.

\begin{figure}[t]
\includegraphics[width=1.0\hsize]{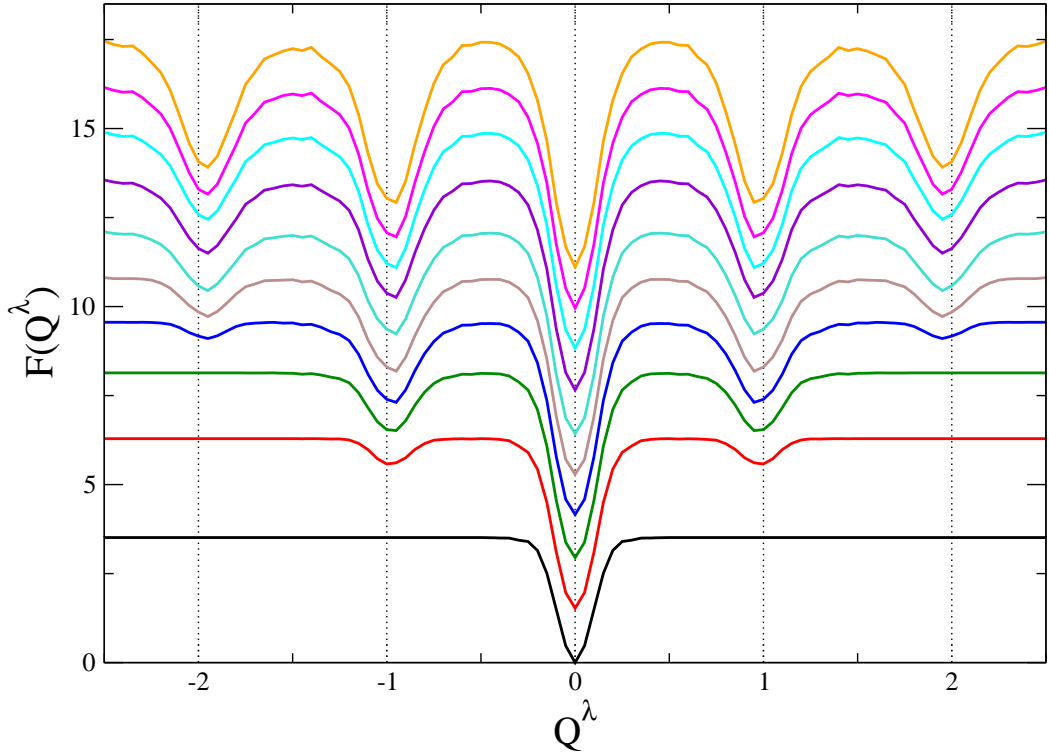}
\caption{\textit{Running average of the metadynamics bias potential
    $V_G(Q,t)$ defined in eq.~(\ref{eq:freeE}) for a run performed
    with $\beta=0.70$ and L=62. The average is taken over 1000 sweeps,
    and shown at intervals of 1000 sweeps, starting from the bottom to
    the top.}}
\label{fig:freenergy}
\end{figure}

\begin{figure}[htp]
  \includegraphics[width=0.46\hsize]{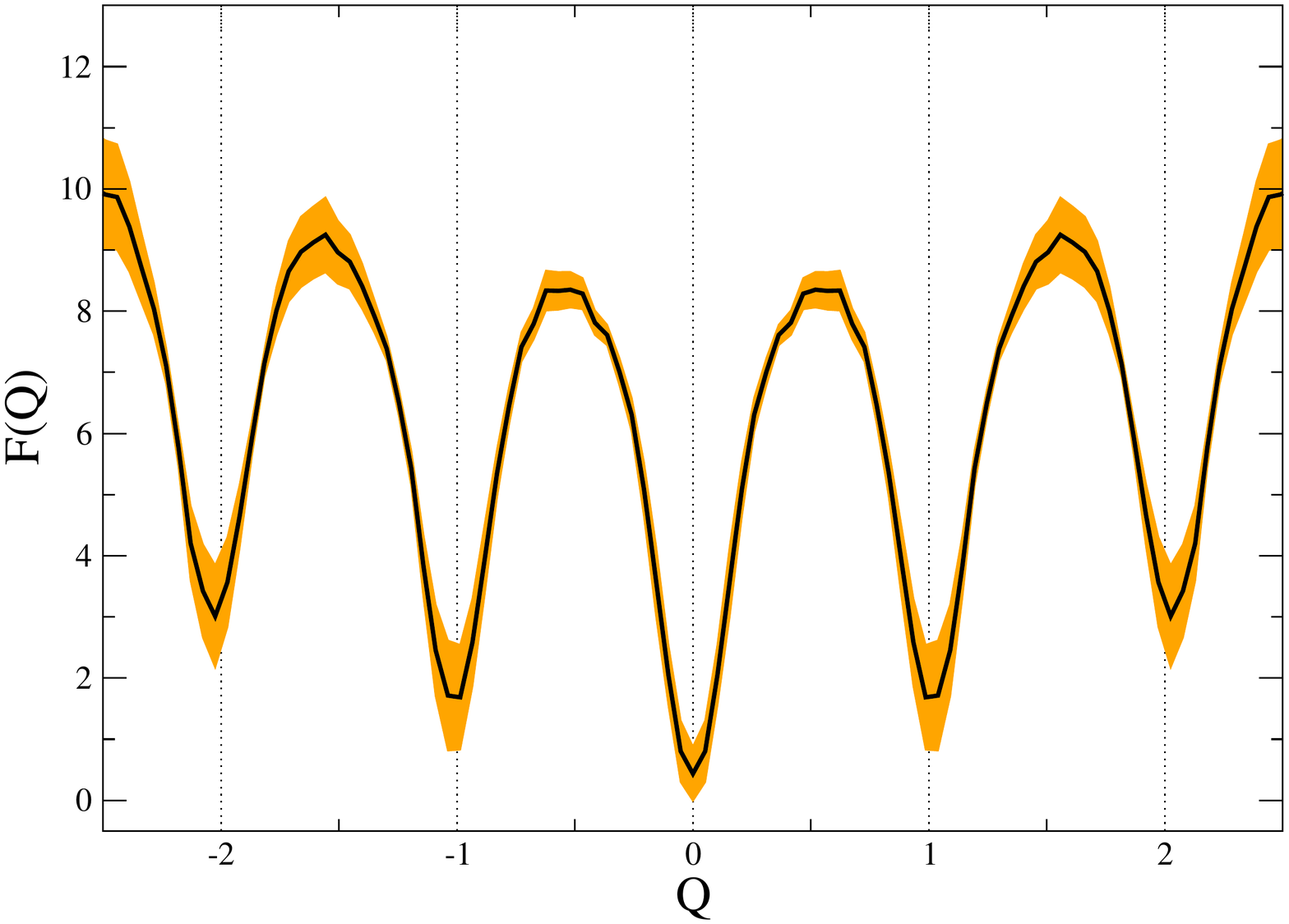}
  \includegraphics[width=0.46\hsize]{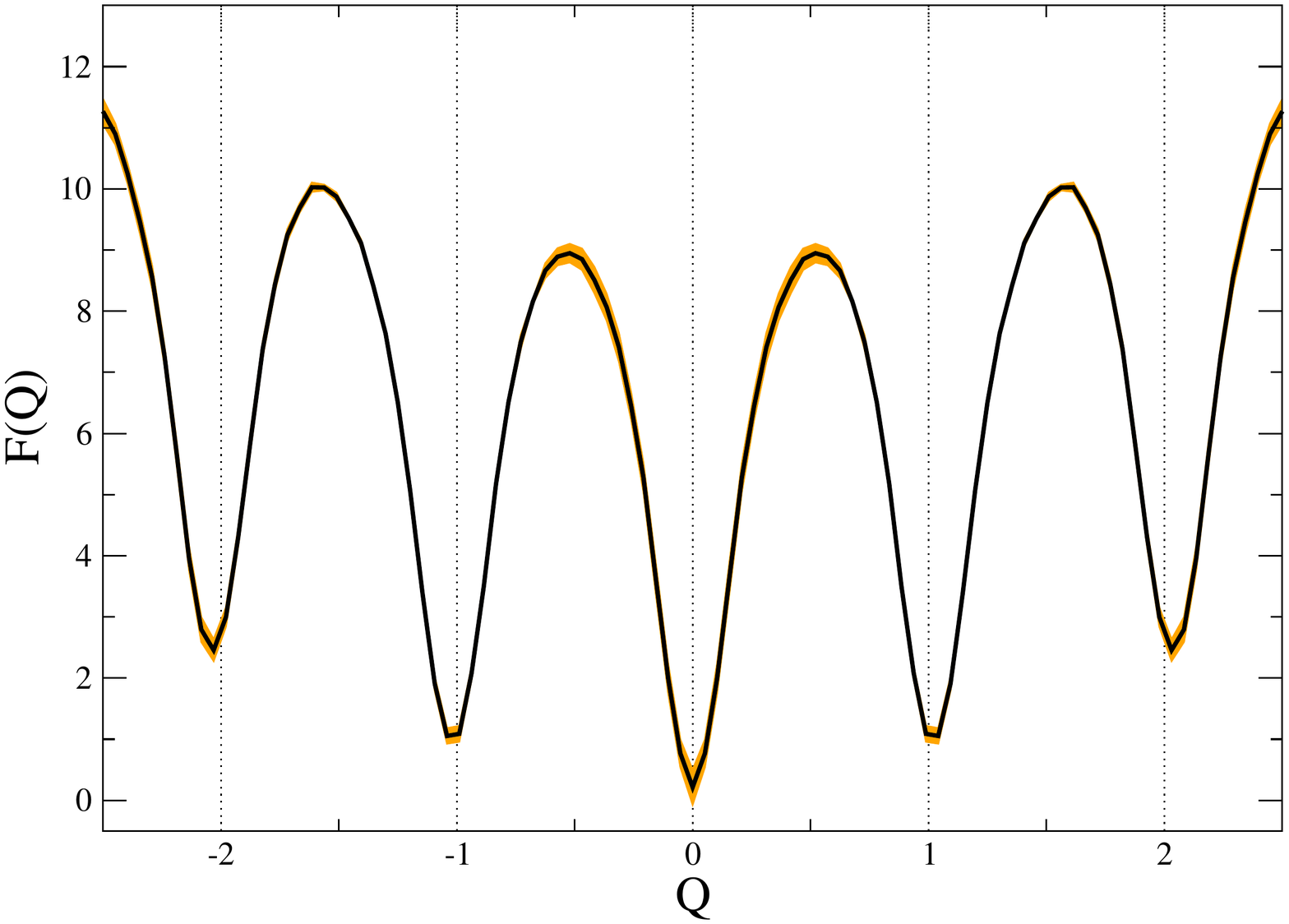}
  \includegraphics[width=0.95\hsize]{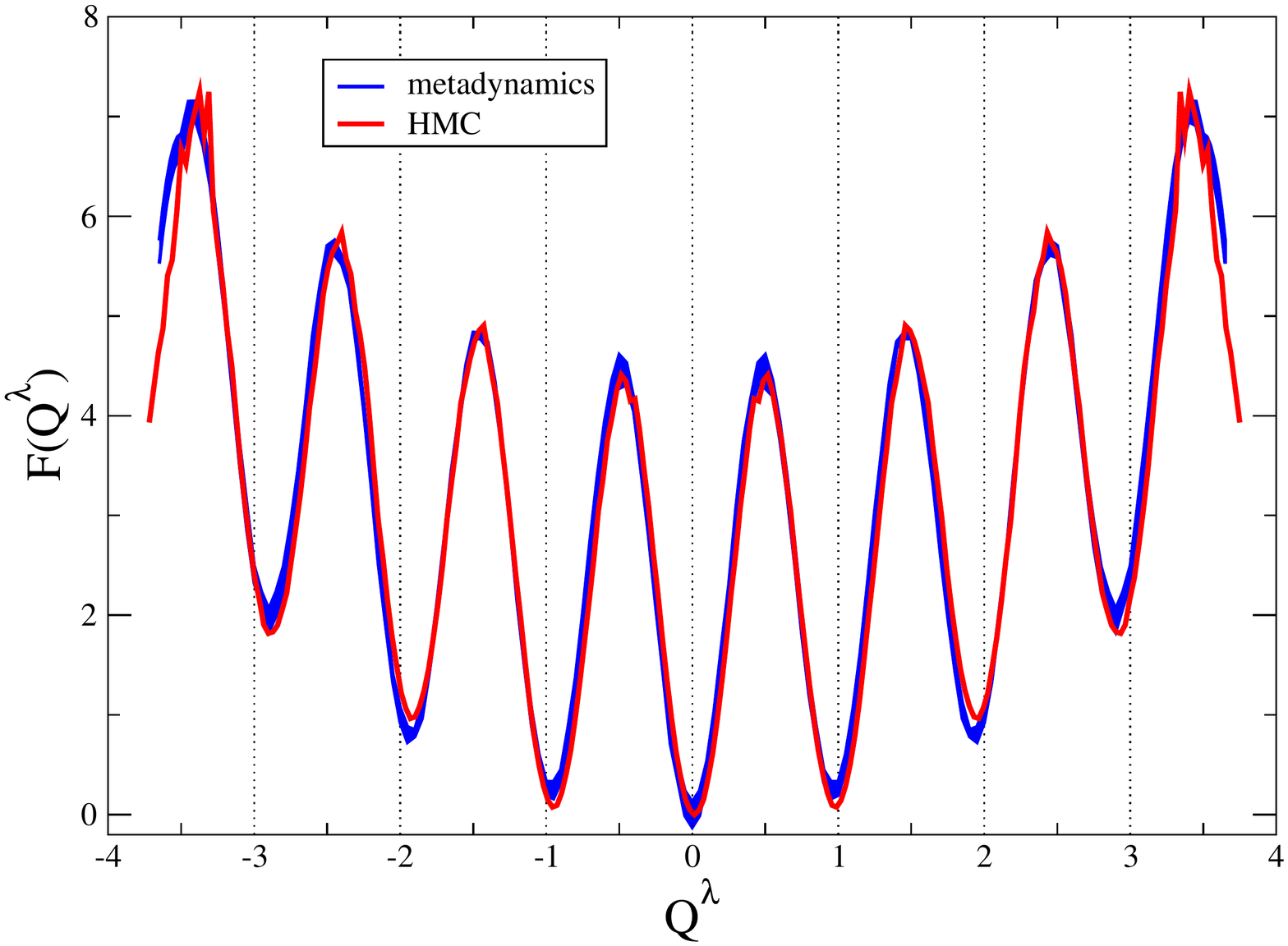}

\caption{\textit{Top: The free energy of the smeared topological
    charge $F\left(Q^{\lambda}\right)$ estimated by metadynamics and
    its statistical uncertainty, shown as a (orange) band at
    $\beta=0.70$ with $L=62$. The average is performed either over 30k
    sweeps (top left) or over 300k sweeps (top right). Bottom:
    comparison between the free energy $F\left(Q^{\lambda}\right)$
    estimated by metadynamics and }$-log(P(Q))$ estimated by
  \textit{the standard HMC. The error on HMC results is not
    shown.}}

\label{fig:freenergyQ30}
\end{figure}

In Fig.~\ref{fig:freenergy} we show the metadynamics bias potential
averaged over progressively larger number of MC sweeps. As the
simulation proceeds, the minima are iteratively filled, until all the
five minima shown in figure are explored, and the bias potential
starts growing evenly in all the $Q$ range. At this point, the free
energy estimator $V_{G}(q,t)$ can be considered converged. Since the
instantaneous value of $V_{G}(q,t)$ is subject to fluctuations, in
order to extract the free energy and its statistical error, we average
$V_{G}(q,t)$ after convergence, namely after 10k HMC sweeps for the
example in Figure. In Fig.~\ref{fig:freenergyQ30} the reconstructed
average free energy of the topological charge
$F\left(Q^{\lambda}\right)$ and its statistical uncertainty, shown as
a (orange) band, estimated by dividing the Monte Carlo history after
equilibration in $n_{p}=4$ different intervals and estimating the
standard deviation of the $n_{P}$ measures. In the bottom panel we
compare the metadynamics result with $-log(P(Q^{\lambda}))$ estimated
in a standard HMC run. Remarkably, the two estimates are fully
consistent within the small error bars, indicating that metadynamics
allows computing reliably the probability distribution of the charge.

\begin{figure}[t]
  \includegraphics[width=0.48\hsize]{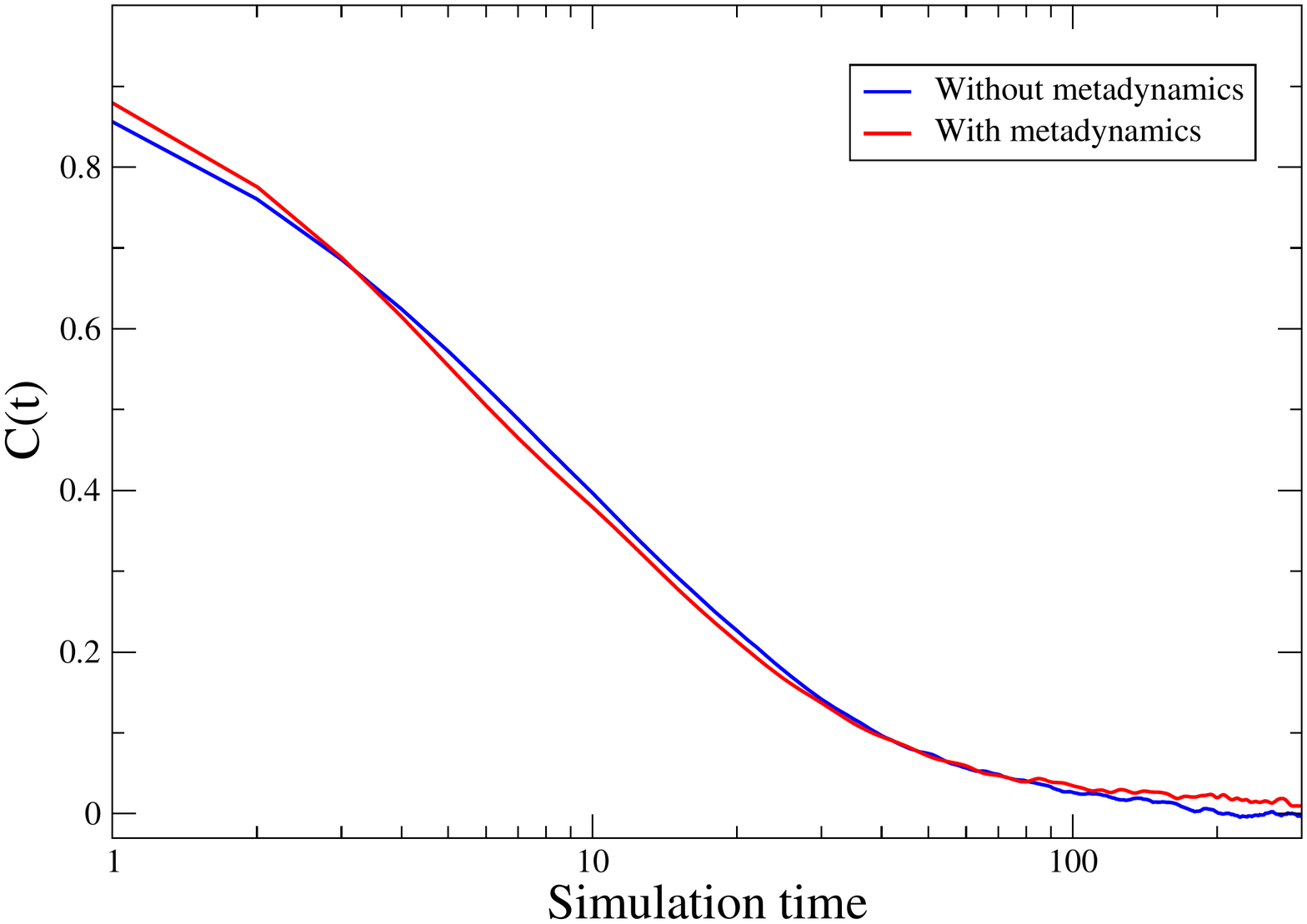}\quad{}
  \includegraphics[width=0.48\hsize]{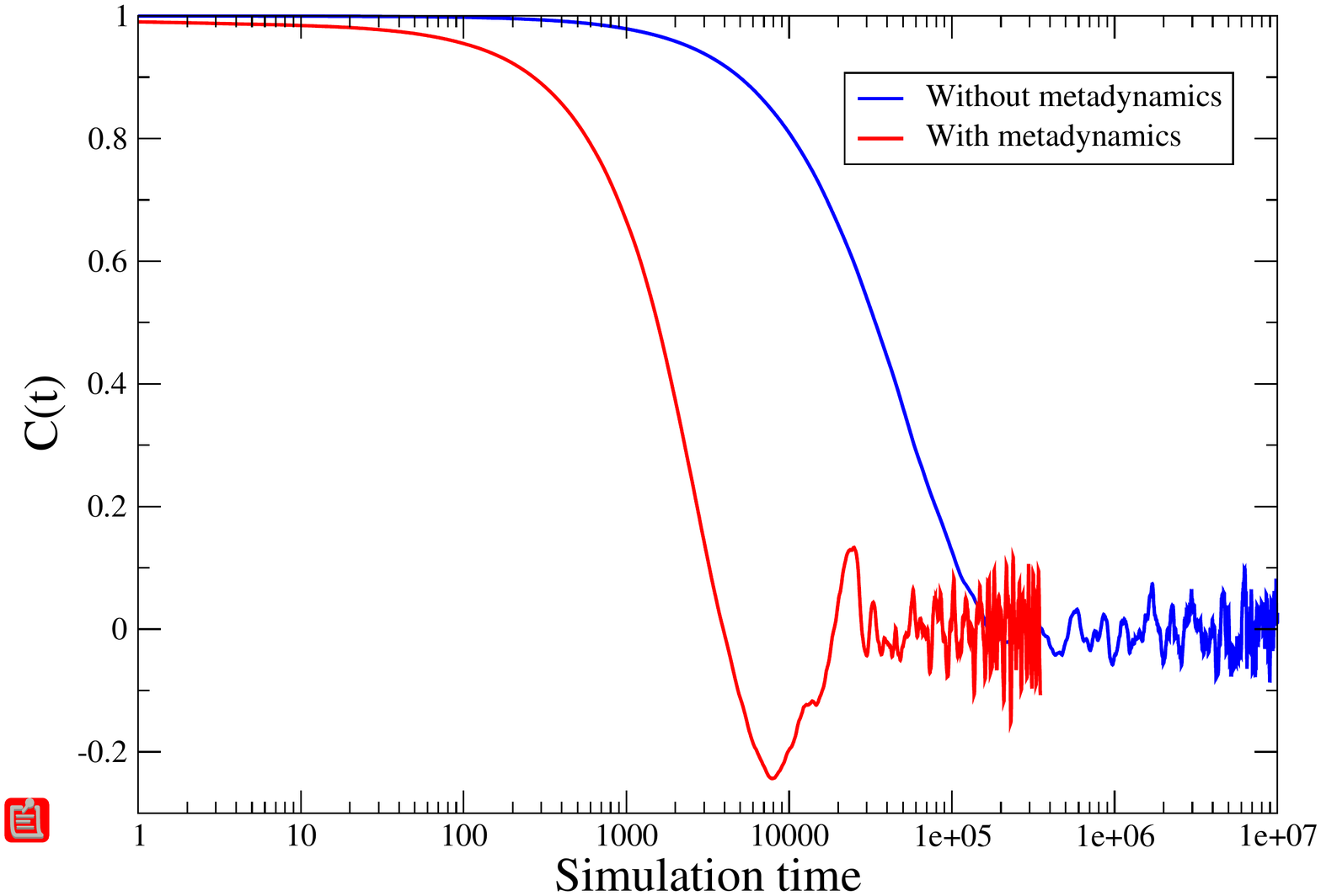}
  \caption{\textit{(a) and (b) Time autocorrelation functions in
      logarithmic x-scale for the magnetization and the topological
      charge. The results correspond to $\beta=0.75$ and $L=60$, for
      which $E=0.6872601(25)$, $\xi_{G}=5.160(3)$ and
      $L/\xi_{G}\sim11.63$.}
  \label{fig:comparison}}
\end{figure}

\begin{figure}[t]
  \includegraphics[width=0.48\hsize]{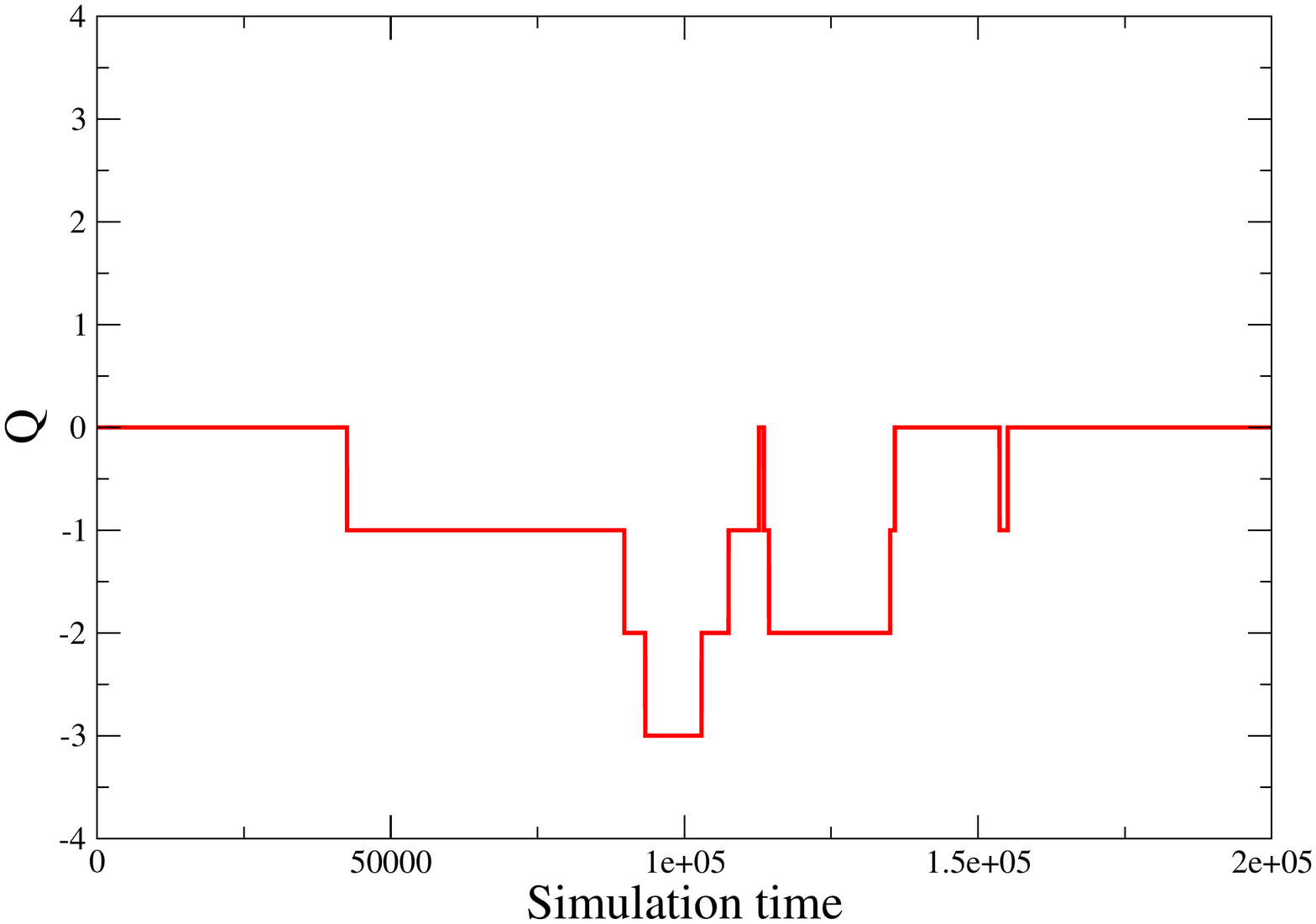}\quad{}
  \includegraphics[width=0.48\hsize]{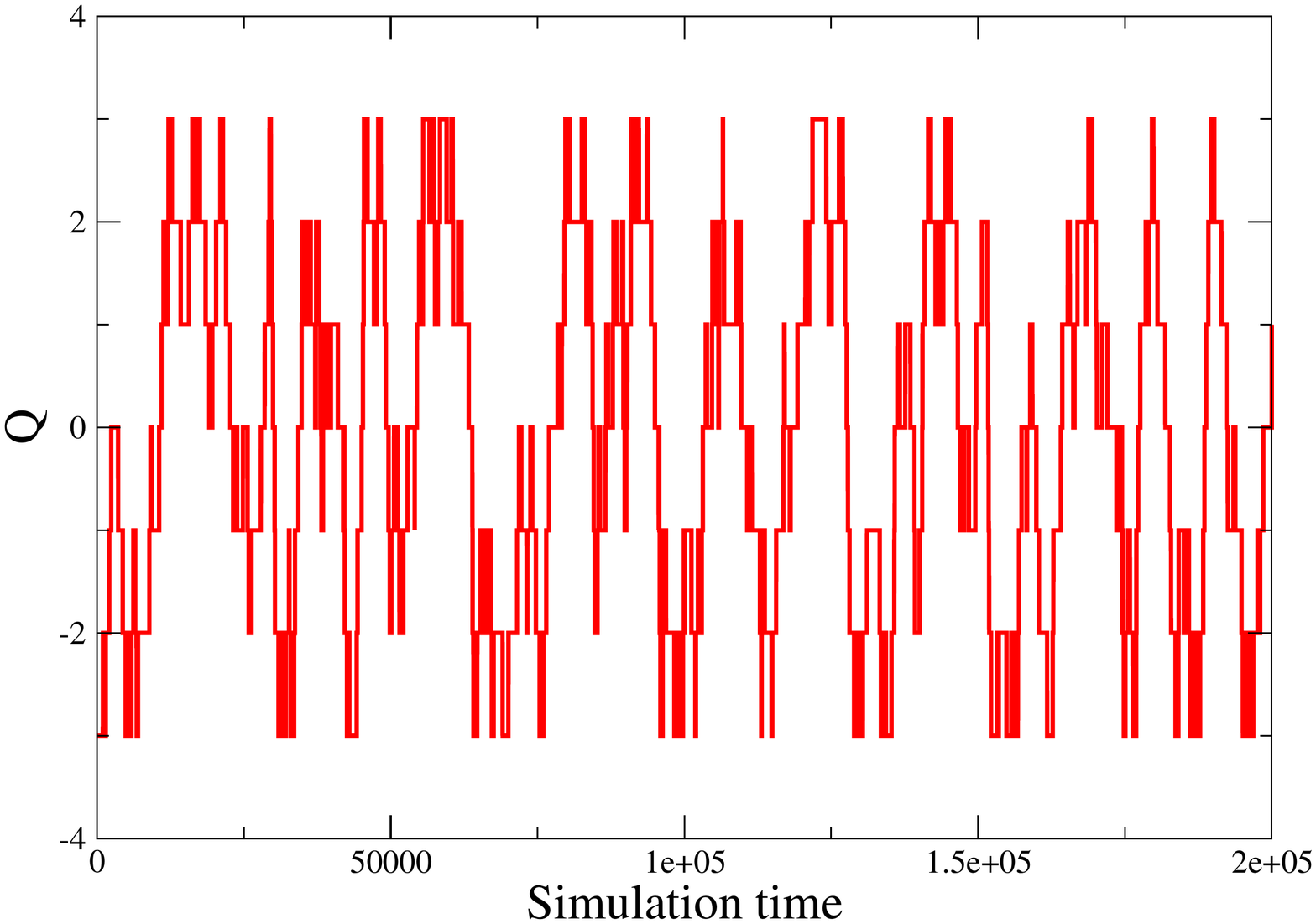}
  \caption{\textit{(a) and (b) Topological charge as a function of the
      number of sweeps with the standard HMC and metadynamics
      respectively at $\beta=0.75$ and $L=60$. In the calculation of
      $\nu$ we have conventionally eliminated jumps in which the
      system returns to the original value of the topological charge
      in less than 10 sweeps.}}
  \label{fig:comparison2}
\end{figure}

The most important effect of the metadynamics bias is reducing the
autocorrelation time of the observables by orders of magnitude. In
Fig.~\ref{fig:comparison} we show the (fictitious) time
autocorrelation function for the magnetization and topological charge
for the standard HMC run and for the metadynamics run. For comparison,
we have chosen for these figures a value of $\beta$ and of the lattice
size $L$ equal to one of the runs presented in
Ref.~\cite{Campostrini:1992ar} (see also \cite{Rossi:1993nz}). The
improvement is significant: by fitting the exponential decrease of the
autocorrelation functions we find $\tau_{Q}=155000\pm10000$ in the HMC
run, whereas the corresponding value for the metadynamics run is
$\tau_{Q}=5600\pm1000$. In both cases we find $\tau_{m}=50\pm10$. A
related quantity is the transition probability per unit time between
two different topological sectors, $\nu$. This quantity is defined as
the number of jumps between two different values of $Q^{g}$ divided by
the total number of sweeps. For the two runs corresponding to
Fig.~\ref{fig:comparison2}, we have $\nu=4.98\cdot10^{-5}$ with the
HMC and $\nu=2.24\cdot10^{-3}$ for metadynamics respectively.

One of the main results of this paper is the reconstructed free
energy, Fig.~\ref{fig:freenergyQ30}, at different values of the
coupling $\beta$ and of the volume. Although the specific form of the
free energy depends on the definition of $Q^{\lambda}$, it contains
some important physical information. $F(Q)$ is very well approximated
by the function
\begin{equation}
  F(Q)=A\, Q^{2}+b\,\sin^{2}(\pi Q),\label{eq:FQfit}
\end{equation}
where $A$ and $b$ are numerical constants which in general,
at fixed $N$, depend on $L$ and $\beta$. 

As an example we fitted the effective potential reported in the
top-left panel of Fig.~\ref{fig:freenergyQ30} relative to $\beta=0.70$
with $L=62$. Taking $c=0$, the results of the fit, in the metadynamics
case, are $A=0.20\pm0.06$ and $B=4.38\pm0.10$, with a $\chi^2/dof~1$.
By using the relation $A\sim1/(2\chi_{t}V)$, this corresponds to
$\chi_{t}=(6.50\pm1.9)\times10^{-4}$ well compatible with the results
given in Tab.~\ref{tab:twob} for $\beta=0.70$ and $L=62$. The
coefficient $b$ is related to the tunneling probability of the system
between different topological sectors. Since most of the transitions
occur with $\Delta Q=\pm1$, we expect that $b\sim\kappa\exp[S_{I}]$,
where $\kappa$ is an entropy factor which increases with the volume
and $S_{I}$ is the one-instanton action $S_{I}=2\pi N\beta$.

\subsection{The central question: Towards the continuum limit}
\label{ssec:scaling}
\begin{figure}[t]
  \includegraphics[width=1.00\hsize]{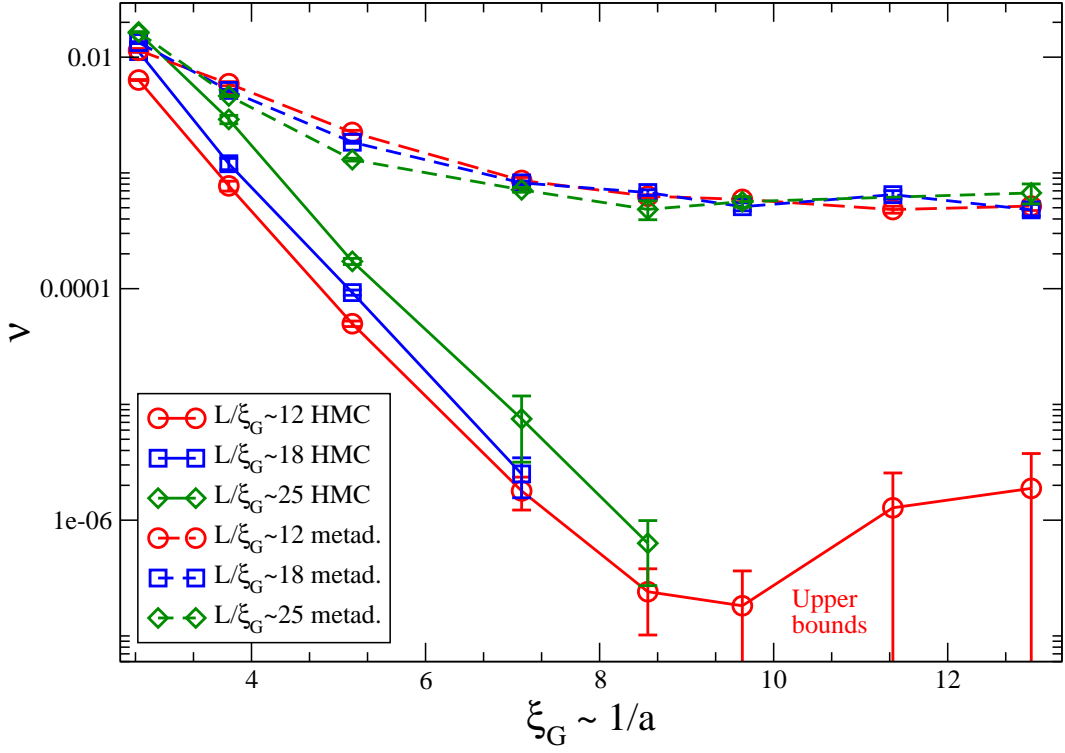}\quad
  \caption{\it Scaling of the frequency $\nu$ as a function of $\xi_G$
    with metadynamics (dashed) and HMC(full line) for three different
    values of $L/\xi_G$.  In the case of the standard HMC, for $a\xi_G
    \gtrsim 9$ the changes of the topological sector are so rare that
    only an upper limit can be estimated.}
    \label{fig:scaling}
\end{figure}

In order to demonstrate that the approach presented in this work
allows addressing efficiently the problem of topological trapping, we
now discuss the scaling of the autocorrelation time as a function of
the lattice spacing $a \sim \xi_G^{-1}$, namely of $\xi_G$, and of the
physical volume $L/\xi_G$. In Fig.~\ref{fig:scaling} and
Tab.~\ref{tab:nu} we display the dependence of $\nu$ on $\xi_G$ with
HMC and metadynamics at several values of $L/\xi_G$.
As expected, because of entropy, in the standard HMC $\nu$ is an
increasing function of the lattice size, and this corresponds to an
increase in the dispersion of the topological charge, namely of
$\langle Q^2\rangle$. At larger values of $\beta$, as we proceed
toward the continuum limit, with the standard HMC, $\nu$ decreases
exponentially as a function of the correlation length
$\xi_G$~\cite{DelDebbio:2002xa,DelDebbio:2004xh}, so that for $a\xi_G
\gtrsim 9$ the changes of the topological sector are so rare that only
an upper limit for $\nu$ can be provided.
\\
The fact that the system is locked in a given topological sector may
severely bias the lattice predictions for several important physical
quantities, {\it in primis} the topological susceptibility. The
apparent small errors with HMC for the topological susceptibility at
large values of $\beta$, for example at $\beta=0.80$, with $L=80$, see
Tab.~\ref{tab:twoa}, are illusory because they refer to an average
taken essentially at fixed topological charge and thus its value is
affected by a systematic error. This is dramatically clear by
comparing the values of the renormalization group invariant product
$\chi_t \xi_G^2$ in the last column of Tab.~\ref{tab:twoa}
and~\ref{tab:twob}. $\beta=0.8$ is just the largest value at which we
could made a comparison between standard HMC and metadynamics, since
above that it is impossible to get sensible results for the
topological susceptibility using HMC. Indeed, with the standard HMC
the number of sweeps necessary to get the correct results increases
exponentially with $\beta$ while in metadynamics increases only
linearly.

The behavior observed with the HMC is to be contrasted with the
results obtained with metadynamics (corrected for the bias using
eq.~\ref{eq:rew}) since in this case $\nu$ is sensibly flatter, and
the simulation spans all the possible sectors of $Q$ allowing to
produce reliable results.

We first analyze the scaling of the performances with the
volume. It is known that in HMC the autocorrelation time of
topological charge does not increase with the volume, as the increased
tunneling rate (due to larger entropy) compensates the larger range of
topological charges to explore. In metadynamics it is a priori not
clear how the mechanism of tunnelling is affected by the volume. We
note empirically (see Fig.~\ref{fig:scaling}) that the tunnelling rate
does not change with the volume, so one would expect that the
algorithm efficiency degrades (mildly) with the volume. This is
partially sustained from the observation of the errors reported on the
last two columns of Tab.~\ref{tab:twob}. Nonetheless we remark that,
no matter which volume is chosen, we still expect metadynamics to
scale better with the lattice spacing than HMC. This means that for
any physical volume there is a critical lattice spacing below which
using metadynamics is convenient.

The final issue that we have to address is the scaling of the error of
observable quantities measured in the metadynamics simulations. As we
proceed to the continuum limit, the weight of configurations with non
integer $Q^\lambda$ decreases, but the system keeps exploring all
values $Q^\lambda\in[-Q^{thr},+Q^{thr}]$ with equal probability after
that the bias potential has converged. The unbiased expectation value
is recovered by the reweighting procedure of eq.~\ref{eq:rew}. The
impact of the reweighting can be estimated modelling the scaling of
the bias potential with the lattice spacing. Assuming that in the
parametrization of eq.~\ref{eq:FQfit} the coefficient $b$, corresponding
to the height of the barriers, grows approximately as $1/a$, one
expects a mild increase of the error due to reweighting. By comparing
Tabs.~\ref{tab:twoa} and ~\ref{tab:twob} one is comforted by observing
that at small $\beta$ (where HMC simulations are reliable) the errors
for $\xi_G$, $E$ and $\chi_m$ obtained with metadynamics are
comparable with those obtained with HMC, and increase very slowly with
$\beta$, not faster than the standard HMC.

For the case of the renormalization invariant combination
$\chi_{t}\xi_{G}^{2}$ we observe that in the case of HMC the errors
growth relatively less than with metadynamics, but the value drops
dramatically at large $\beta$ due to the freezing of the topological
charge. With metadynamics we find similar values across the whole
range of explored lattice spacings, signaling that the algorithm
samples correctly the distribution of the topological charge, as
opposed to HMC.

In spite of the great effort undertaken in the
past~\cite{DelDebbio:2002xa, DelDebbio:2004xh, Flynn:2015uma} even for
the HMC algorithm it is not clear yet whether the topological modes
are affected by exponential or power-like slowing down. In this first
work we do not aim at determining accurately the scaling properties of
our current implementation of metadynamics. We can however note that
the observed growth of the error on the topological susceptibility as
a function of the lattice spacing is compatible with a power law
divergence of topological autocorrelation time, with an exponent that
we estimate to be $2\div 3$. This has to be compared with the pure HMC
simulations, for which the exponent of the power-like ansatz has
steadily been suggested to be close to 5. We consider this a
substantial improvement, that allows already to have reliable results
at $\beta$ much larger that the HMC. Based on previous
experience~\cite{bias_exchange}, further improvement might be obtained
including a larger set of collective variables to the algorithm,
possibly coupling including in the bias other slow modes.

\begin{minipage}{\textwidth}
  { \renewcommand\arraystretch{0.8}
    \begin{center}
      \begin{longtable}[t]{|c|c|c|c|c|c|c|c|c|}
        \hline 
        $\beta=\frac{1}{gN}$ & $N_{s}$ & $L$ & $\xi_{G}$ & $L/\xi_{G}$ & $E$ & $\chi_{m}\times 10^4$ & $\chi_{t}\times 10^4$ & $\chi_{t}\xi_{G}^{2}\times 10^4$\\
        \hline 
        \multirow{3}{*}{0.65}
        & 3M & 32 & 2.7009(13) & 11.848(15) & 0.799324(9) & 12.182(3) & 11.66(18) & 85.1(1.3)\\
        & 3M & 46 & 2.7024(19) & 17.022(12) & 0.799352(8) & 12.148(3) & 11.91(26) & 87(2)\\
        & 3M & 64 & 2.7075(23) & 23.64(2)   & 0.799357(4) & 12.147(2) & 11.6(3)   & 84.9(2.2)\\
        \hline 
        \multirow{3}{*}{0.70}
        & 3M & 44 & 3.748(3) & 11.739(8)  & 0.739053(4) & 19.613(8) & 5.56(24) & 78(3)\\
        & 3M & 62 & 3.742(3) & 16.57(5)   & 0.739072(4) & 19.545(4) & 5.93(14) & 83(2)\\
        & 3M & 88 & 3.743(4) & 23.508(27) & 0.739066(3) & 19.543(5) & 5.88(26) & 82(4)\\
        \hline 
        \multirow{3}{*}{0.75}
        & 3M & 60  & 5.169(7)  & 11.608(16) & 0.687256(6)   & 31.890(31) & 2.5(3) & 67(8)\\
        & 3M & 84  & 5.154(8)  & 16.299(24) & 0.687270(6)   & 31.725(21) & 3.7(5)   & 97(13)\\
        & 3M & 120 & 5.152(16) & 23.28(7)   & 0.6872678(21) & 31.723(18) & 2.9(5) & 78(13)\\
        \hline 
        \multirow{3}{*}{0.80}
        & 3M & 80  & 7.141(13) & 11.203(20) & 0.642261(4)   & 52.48(6)   & 0.74(12) & 38(6)\\
        & 3M & 114 & 7.102(8)  & 16.051(18) & 0.6422752(22) & 52.120(23) & 0.29(6)  & 15(3)\\
        & 3M & 160 & 7.075(20) & 22.61(6)   & 0.6422784(15) & 52.044(22) & 0.25(8)  & 13(4)\\ 

        \hline 
        \caption{{\it Parameters and measured quantities of the standard HMC
            simulation. All the quantities are measured every 5 sweeps except
            the energy and the topological charge which are measured every
            sweep.}}
        \label{tab:twoa}
      \end{longtable}

      \begin{longtable}[tb]{|c|c|c|c|c|c|c|c|c|}
        \hline 
        $\beta=\frac{1}{gN}$ & $N_{s}$ & $L$ & $\xi_{G}$ & $L/\xi_{G}$ & $E$ & $\chi_{m}\times 10^4$ & $\chi_{t}\times 10^4$ & $\chi_{t}\xi_{G}^{2}\times 10^4$\\
        \hline 
        \multirow{3}{*}{0.65}
        & 3M & 32 & 2.7000(20) & 11.852(9)  & 0.799310(14) & 12.181(4)   & 11.44(14) & 83.4(1.1)\\
        & 3M & 46 & 2.6994(24) & 17.041(15) & 0.799366(6)  & 12.1448(29) & 11.83(19) & 86.2(1.4)\\
        & 3M & 64 & 2.702(4)   & 23.696(36) & 0.799354(6)  & 12.146(3)   & 11.32(15) & 82.6(1.1)\\
        \hline 
        \multirow{3}{*}{0.70} 
        & 3M & 44 & 3.743(4) & 11.756(12) & 0.739049(10) & 19.608(10) & 5.68(11) & 79.6(1.5)\\
        & 3M & 62 & 3.742(6) & 16.576(16) & 0.739069(6)  & 19.550(7)  & 5.66(12) & 79.2(1.7)\\
        & 3M & 88 & 3.750(5) & 23.47(3)   & 0.739071(4)  & 19.539(4)  & 5.68(25) & 80(3)\\
        \hline 
        \multirow{3}{*}{0.75} 
        & 3M & 60 & 5.149(7)  & 11.653(15) & 0.687281(7) & 31.795(21) & 3.01(7)  & 79.8(1.8)\\
        & 3M & 84 & 5.141(7)  & 16.339(23) & 0.687270(5) & 31.717(13) & 2.91(14) & 77(3)\\
        & 3M & 120& 5.183(11) & 23.15(5)   & 0.687272(4) & 31.735(10) & 3.2(3)   & 86(8)\\
        \hline 
        \multirow{3}{*}{0.80}
        & 3M & 80  & 7.079(7)  & 11.302(11) & 0.642289(5)   & 52.141(30) & 1.78(19) & 87(10)\\
        & 3M & 114 & 7.098(19) & 16.06(4)   & 0.642288(4)   & 51.98(4)   & 1.80(23) & 89(12)\\
        & 3M & 160 & 7.072(19) & 22.62(6)   & 0.6422797(26) & 51.971(22) & 1.79(20) & 88(10)\\
        \hline 
        0.85  & 3M & 108 & 9.672(18)  & 11.166(21) & 0.602879(5)  & 85.91(8)    & 1.00(16) & 91(15)\\
        \hline 
        0.875 & 3M & 126 & 11.073(21) & 11.379(21) & 0.584949(4)  & 110.72(14)  & 0.69(16) & 77(20)\\
        \hline 
        0.9   & 3M & 148 & 13.32(4)   & 11.112(30) & 0.5680742(29) & 143.28(22) & 0.42(23) & 73(41)\\
        \hline 
        \caption{{\it Parameters and measured quantities of the
            metadynamics simulation.}}
        \label{tab:twob}
      \end{longtable}
  \end{center}}
\end{minipage}

\begin{table}[htp]
\begin{center}
\begin{tabular}{|c|c|c|c|} \hline 
$\beta$ & $L$ & $\nu$ HMC & $\nu$ metadynamics\\ \hline
0.65 & 12 & $(6.4 	\pm 0.1) \times  10^{-3}$ &  $(11.5	\pm 0.4) \times  10^{-3}$\\ 
0.70 & 12 & $(7.7 	\pm 0.8) \times  10^{-4}$ &  $(5.9	\pm 0.2) \times  10^{-3}$\\ 
0.75 & 12 & $(5.0 	\pm 0.3) \times  10^{-5}$ &  $(2.2	\pm 0.1) \times  10^{-3}$\\ 
0.80 & 12 & $(1.8 	\pm 0.6) \times  10^{-6}$ &  $(8.7	\pm 0.5) \times  10^{-4}$\\ 
0.83 & 12 & $(2.4 	\pm 1.4) \times  10^{-7}$ &  $(6.2	\pm 0.2) \times  10^{-4}$\\ 
\hline
0.65 & 18 & $(11.1	\pm 0.1) \times  10^{-3}$ &  $(13.4	\pm 0.2) \times  10^{-3}$\\ 
0.70 & 18 & $(1.1 	\pm 0.1) \times  10^{-3}$ &  $(5.2	\pm 0.1) \times  10^{-3}$\\ 
0.75 & 18 & $(9.3 	\pm 0.5) \times  10^{-5}$ &  $(1.84	\pm 0.04) \times  10^{-3}$\\ 
0.80 & 18 & $(2.5	\pm 0.9) \times  10^{-6}$ &  $(8.2	\pm 0.3) \times  10^{-4}$\\ 
\hline
0.65 & 25 & $(16.2	\pm 0.2) \times  10^{-3}$ &  $(16.4	\pm 0.2) \times  10^{-3}$\\ 
0.70 & 25 & $(2.9	\pm 0.2) \times  10^{-3}$ &  $(4.62	\pm 0.06) \times  10^{-3}$\\ 
0.75 & 25 & $(1.7 	\pm 0.1) \times  10^{-4}$ &  $(1.30	\pm 0.04) \times  10^{-3}$\\ 
0.80 & 25 & $(7.5	\pm 4.3) \times  10^{-6}$ &  $(7.1	\pm 0.3) \times  10^{-4}$\\
\hline
\end{tabular}
\caption{{\it Values of the frequency $\nu$ as a function of the
    coupling for different physical volumes. In the table we only give
    the values of $\nu$ that have been reliably determined, excluding
    those corresponding to $\beta$ so large that, in the case of HMC,
    only an upper bound can be given. }}
\end{center}
\label{tab:nu}
\end{table}

\section{Conclusions}
\label{sec:conclu}

In this paper we have shown that the {\it metadynamics}
approach~\cite{laio2002escaping,laio2} can be used to simulate
$CP^{N-1}$ improving dramatically the problem of the slowing down
observed in numerical simulations for quantities related to the
topological charge. In particular we have studied the $N=21$ case,
showing that we are able to reconstruct the free energy of the
topological charge, $F(Q)$, as a function of the coupling and of the
volume.

The much reduced slowing down allows us to study a range of $\beta$
much larger than that available with ordinary HMC. Further improvement
might be obtained by biasing a larger set of collective variables.

It seems straightforward to extend the general method exposed in this
paper to the case of QCD. It will be interesting to investigate
whether metadynamics will also work in this case.

The possibility of measuring of $F(Q)$ suggests that we can further
improve the efficiency of the algorithm by using the information about
its form as obtained in a preliminary run,
cfr. eq.~(\ref{eq:FQfit}). This would allow simulating our system by
using as importance sampling the knowledge of free energy itself added
to the original action of our theory.

The knowledge of the free energy of the topological charge
($Q^{\lambda}$ in our case) is very important for the study of
relevant physical quantities. Actually, if we know the free energy of
the topological charge, then we can compute the expectation value of
any physical quantity at arbitrary values of $\theta$-vacuum, and not
only close to the origin, by simply averaging them with the
appropriate weight $\sim\cos(\theta Q)\, e^{-F(Q)}$, where we used the
property that $F(-Q)=F(Q)$. Metadynamics might thus also offer a
solution to the difficulties encountered in simulating theories with
complex actions (for example QCD with a non zero $\theta$ term or at
finite chemical potential).

\appendix
\section{}\label{app:SmeU1}
Taking inspiration from QCD we define recursively the
$\left(n+1\right)$-level stout smeared link
$\lambda_{m;\mu}^{\left(n+1\right)}$ in terms of the $n$-level stout
smeared link $\lambda_{m;\mu}^{n}$ as
\[
\lambda_{m;\mu}^{\left(n+1\right)}=e^{iQ_{m;\mu}^{\left(n\right)}}\lambda_{m;\mu}^{\left(n\right)}\,,
\]
where 
\[
Q_{m;\mu}^{\left(n\right)}={\rm Im}\left(\rho
  S_{m;\mu}^{\left(n\right)}\lambda^{\left(n\right)}_{m;\mu}\right)\,,
\]
 and $S_{m;\mu}^{\left(n\right)}$ are the \emph{staples} %
\begin{minipage}[c]{0.12\columnwidth}%
  \includegraphics[width=\columnwidth]{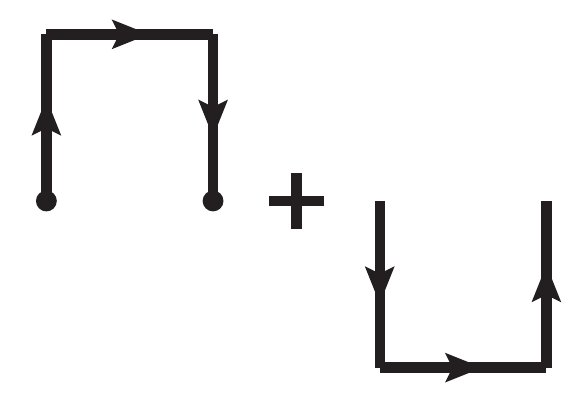}%
\end{minipage}:
\[
S^{\left(n\right)}_{m;\mu}=\sum_{v\neq\mu}\left[\lambda^{\left(n\right)}_{m;\nu}\lambda^{\left(n\right)}_{m+\hat{\nu};\mu}\lambda^{\left(n\right)
    *}_{m+\hat{\mu};\nu}+\lambda^{\left(n\right)*}_{m-\hat{\nu};\nu}\lambda^{\left(n\right)}_{m-\hat{\nu};\mu}\lambda^{\left(n\right)}_{m+\hat{\nu}+\hat{\mu};\nu}\right]\,,
\]
being $\rho$ a small real number. The recursive procedure is based on
the $0$-level stout smeared links, that are nothing but the original
(non-smeared) ones \includegraphics[width=0.06\columnwidth]{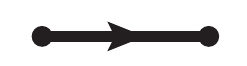}.

As an effect of the exponentiation, the level-1 link results to be the
average of the original link and the collection of all the infinite
paths surrounding the two nearby $1\times1$ plaquettes:

\begin{center}
  \includegraphics[width=1cm]{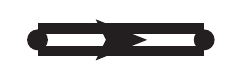}
  $\boldsymbol{=}$ \includegraphics[width=1cm]{link}
  $\boldsymbol{+}${\Large{$\frac{\rho}{2}$}} $\Big\{$
  \raisebox{-0.4\height}{\includegraphics[width=6cm]{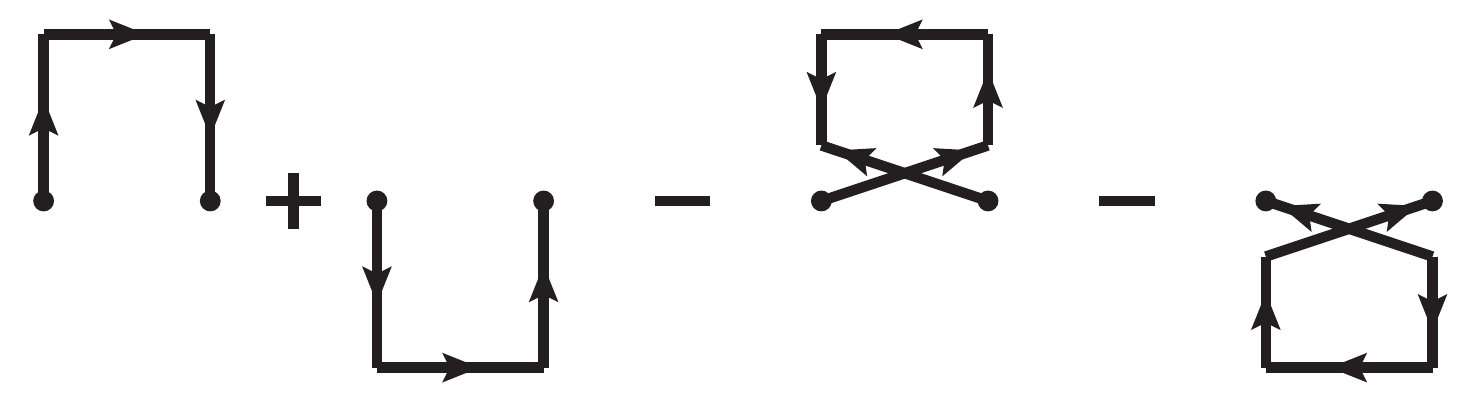}}
  $\boldsymbol{+}$ \dots $\Big\}$\,.
\end{center}%

The definition is recursive and the $n$-level smeared link will
involve contribution of links as far as $d\leq n$ sites (each of them
damped by a power $\rho^{d}$). Averaging a link with paths built in
terms of its neighbors suppresses ultraviolet fluctuations, and
reduces the noise in observables built in terms of the links. The
parameters $\rho$ and $n$ can be tuned separately, but what indeed
matters is the product $\rho\, n$. We found it convenient to use $n=2$
and vary linearly $\rho$ as a function of $\beta$ ($\rho = 0.08$ at $
\beta = 0.65$ and $\rho=0.13$ at $\beta=0.90$) which allows to
separate the distribution probabilities of $Q^\lambda$ for different
topological sectors as explained in Sec.~\ref{sec:meta}.

\section*{Acknowledgments}

We are particularly grateful to J.~Flynn, A.~Juttner, M.~Testa and
E.~Vicari for several useful discussions.  This work was partially supported  by
the ERC-2010 DaMESyFla Grant Agreement Number: 267985, and by the MIUR
(Italy) under a contract PRIN10-11 protocollo 2010YJ2NYW.  FS received
funding from the European Research Council under the European
Community Seventh Framework Programme (FP7/2007-2013) ERC grant
agreement No 279757.

All calculations were performed on the Ulysses HPC facility maintained
by SISSA - ITCS (Information Technology and Computing Services). We
acknowledge Antonio Lanza, Piero Calucci and all members of ITCS for
their continuous support and availability.


\begin{thebibliography}{99}

\bibitem{Rossi:1993nz}
  P.~Rossi and E.~Vicari,
  Phys.\ Rev.\ D {\bf 48} (1993) 3869
  [hep-lat/9301008].

\bibitem{DelDebbio:2002xa} 
  L.~Del Debbio, H.~Panagopoulos and E.~Vicari, JHEP {\bf 0208} (2002)
  044 [hep-th/0204125].

\bibitem{DelDebbio:2004xh}
  L.~Del Debbio, G.~M.~Manca and E.~Vicari, Phys.\ Lett.\ B {\bf 594}
  (2004) 315 [hep-lat/0403001].

\bibitem{Luscher:2010we}
  M.~Luscher, PoS LATTICE {\bf 2010} (2010) 015 [arXiv:1009.5877
  [hep-lat]].

\bibitem{Schaefer:2010hu} 
  S.~Schaefer {\it et al.} [ALPHA Collaboration], Nucl.\ Phys.\ B {\bf
    845} (2011) 93 [arXiv:1009.5228 [hep-lat]].

\bibitem{Luscher:2011kk} 
  M.~Luscher and S.~Schaefer, JHEP {\bf 1107} (2011) 036
  [arXiv:1105.4749 [hep-lat]].

\bibitem{Chowdhury:2013mea} 
  A.~Chowdhury, A.~Harindranath, J.~Maiti and P.~Majumdar, JHEP {\bf
    1402} (2014) 045 [arXiv:1311.6599 [hep-lat]].

\bibitem{Brower:2014bqa} 
  R.~C.~Brower {\it et al.} [LSD Collaboration], Phys.\ Rev.\ D {\bf
    90} (2014) 1, 014503 [arXiv:1403.2761 [hep-lat]].

\bibitem{Flynn:2015uma}
  J.~Flynn, A.~Juttner, A.~Lawson and F.~Sanfilippo, arXiv:1504.06292
  [hep-lat].

\bibitem{Engel:2011re}
  G.~P.~Engel and S.~Schaefer, Comput.\ Phys.\ Commun.\ {\bf 182}
  (2011) 2107 [arXiv:1102.1852 [hep-lat]].

\bibitem{Veneziano}
  G. Veneziano, Nucl. Phys.  B159 , 213 (1979).

\bibitem{Witten}
  E. Witten, Nucl. Phys.B156, 269 (1979).

\bibitem{DelDebbio:2004ns}
  L.~Del Debbio, L.~Giusti and C.~Pica, Phys.\ Rev.\ Lett.\ {\bf 94}
  (2005) 032003 [hep-th/0407052].

\bibitem{Ce:2015qha} 
  M.~Cè, C.~Consonni, G.~P.~Engel and L.~Giusti,
  Phys.\ Rev.\ D {\bf 92}, no. 7, 074502 (2015)
  [arXiv:1506.06052 [hep-lat]].

\bibitem{Shore:2007yn} 
  G.~M.~Shore, Lect.\ Notes Phys.\ {\bf 737} (2008) 235
  [hep-ph/0701171].

\bibitem{Aidala:2012mv}
  C.~A.~Aidala, S.~D.~Bass, D.~Hasch and G.~K.~Mallot, Rev.\ Mod.\
  Phys.\ {\bf 85} (2013) 655 [arXiv:1209.2803 [hep-ph]].

\bibitem{Marchand}
  C.~Marchand, DOI: 10.3204/DESY-PROC-2012-02/20 Conference:
  C12-03-26.1, p.27-36.

\bibitem{Mulders}
  J. Mulders, Prog.\ Part.\ Nucl.\ Phys.\ {\bf 55}, 243 (2005).  
 
\bibitem{deForcrand:1997fm}
  P.~de Forcrand, J.~E.~Hetrick, T.~Takaishi and A.~J.~van der Sijs,
  Nucl.\ Phys.\ Proc.\ Suppl.\  {\bf 63} (1998) 679
  [hep-lat/9709104].

\bibitem{Ramos:2012bb} 
  A.~Ramos,
  PoS LATTICE {\bf 2012}, 193 (2012)
  [arXiv:1212.3800 [hep-lat]].

\bibitem{McGlynn:2013ava} 
  G.~McGlynn and R.~D.~Mawhinney,
  PoS Lattice {\bf 2013}, 027 (2014)
  [arXiv:1311.3695 [hep-lat]].

\bibitem{Gerber:2014bia} 
  U.~Gerber, I.~Bautista, W.~Bietenholz, H.~Mejía-Díaz and C.~P.~Hofmann,
  PoS LATTICE {\bf 2014}, 320 (2014)
  [arXiv:1410.0426 [hep-lat]].

\bibitem{Endres:2015yca} 
  M.~G.~Endres, R.~C.~Brower, W.~Detmold, K.~Orginos and A.~V.~Pochinsky,
  arXiv:1510.04675 [hep-lat].

\bibitem{Gambhir:2015rha} 
  A.~S.~Gambhir and K.~Orginos,
  PoS LATTICE {\bf 2014}, 043 (2015)
  [arXiv:1506.06118 [hep-lat]].

\bibitem{Dromard:2015nba} 
  A.~Dromard, W.~Bietenholz, U.~Gerber, H.~MejÃ­a-DÃ­az and M.~Wagner,
  arXiv:1510.08809 [hep-lat].

\bibitem{laio2002escaping}
  A.~Laio, M.~Parrinello, Proceedings of the National Academy of
  Sciences, 2002, 99.20: 12562-12566.

\bibitem{laio2} 
  A.~Laio and F.L.~Gervasio, Rep. Prog. Phys. 71, 126601, 2008

\bibitem{wang2001}
  F.~Wang, F. and D.~P.~Landau Phys.\ Rev.\ Lett.\ {\bf 86} (2001)
  2050.  [arXiv:cond-mat/0011174]

\bibitem{laiostar}
  Y. Crespo, F. Marinelli, F. Pietrucci, A. Laio, Physical Review E,
  81, 055701 (2010).

\bibitem{DiVecchia:1981eh}
  P.~Di Vecchia, A.~Holtkamp, R.~Musto, F.~Nicodemi and R.~Pettorino,
  Nucl.\ Phys.\ B {\bf 190} (1981) 719.
  
\bibitem{Berg:1981er}
  B.~Berg and M.~Luscher, Nucl.\ Phys.\ B {\bf 190} (1981) 412.

\bibitem{Campostrini:1992ar}
  M.~Campostrini, P.~Rossi and E.~Vicari, Phys.\ Rev.\ D {\bf 46}
  (1992) 2647.

\bibitem{bias_exchange}
  S.~Piana, A.~Laio, A bias-exchange approach to
  protein folding Journal of Physical Chemistry B, 111, 4553 (2007).
  
\bibitem{Bonati:2014tqa}
  C.~Bonati and M.~D'Elia, Phys.\ Rev.\ D {\bf 89} (2014) 105005
  [arXiv:1401.2441 [hep-lat]].

\bibitem{Albanese:1987ds}
  M.~Albanese {\it et al.}  [APE Collaboration], Phys.\ Lett.\ B {\bf
    192} (1987) 163.

\bibitem{Hasenfratz:2001hp}
  A.~Hasenfratz and F.~Knechtli, Phys.\ Rev.\ D {\bf 64} (2001) 034504
  [hep-lat/0103029].

\bibitem{Morningstar:2003gk}
  C.~Morningstar and M.~J.~Peardon, Phys.\ Rev.\ D {\bf 69} (2004)
  054501 [hep-lat/0311018].


\end{thebibliography}
\end{document}